\documentclass[preprint]{aastex}

\newcommand{\ls}
 {\mathrel{\hbox{\rlap{\hbox{\lower4pt\hbox{$\sim$}}}\hbox{$<$}}}}
\newcommand{\gs}
 {\mathrel{\hbox{\rlap{\hbox{\lower4pt\hbox{$\sim$}}}\hbox{$>$}}}}
\newcommand{\degg}{\hbox{$^\circ$}}

\newcommand{\arcs}{\hbox{$^{\prime\prime}$}}
\newcommand{\et}{et al.\ }
\newcommand{\ginga}{{\it Ginga}}
\newcommand{\rosat}{{\it ROSAT}}
\newcommand{\asca}{{\it ASCA}}
\newcommand{\xmm}{{\it XMM-Newton}}
\newcommand{\chandra}{{\it Chandra}}

\newcommand{\ariel}{{\it Ariel V}}

\newcommand{\exosat}{{\it EXOSAT}}

\newcommand{\mrk}{{Mkn 509}}

\def\la{\mathrel{\hbox{\rlap{\hbox{\lower4pt\hbox{$\sim$}}}{\raise2pt\hbox{$<$}}
}}}
\def\ga{\mathrel{\hbox{\rlap{\hbox{\lower4pt\hbox{$\sim$}}}{\raise2pt\hbox{$>$}}
}}}

\begin{document}

\title{An XMM-Newton observation of the luminous Seyfert 1 galaxy
Markarian 509}
\shorttitle{AN XMM-Newton observation of Mkn 509}
\shortauthors{Pounds \et}
\author{ Ken Pounds\altaffilmark{1},
	James Reeves\altaffilmark{1},
	Paul O'Brien\altaffilmark{1},
	Kim Page\altaffilmark{1}, 
        Martin Turner\altaffilmark{1},
        \& Sergei Nayakshin\altaffilmark{2}}

%\email{kap@star.le.ac.uk}

\altaffiltext{1}{X-ray Astronomy Group; University of
Leicester; Leicester
LE1 7RH; United Kingdom}

\altaffiltext{2} {USRA and NASA/Goddard Space Flight Center, Code 661, 
Greenbelt, MD, 20771, USA}

\begin{abstract}

We present the spectral analysis of an early \xmm\ observation 
of the luminous Seyfert 1 galaxy \mrk.
We find the hard (2--10~keV) continuum slope, including reflection, 
to be somewhat flatter ($\Gamma=1.75$) than for
a typical BLS1. The most obvious feature in the hard X-ray spectrum is a
narrow emission line near 6.4~keV, with an equivalent width of $\sim$50 eV.
The energy and strength of this line is consistent with 
fluorescence from `neutral' iron in the molecular torus, and we note
the emerging ubiquity of this feature in \xmm\ and \chandra\ observations 
of Seyfert 1 galaxies over a wide luminosity range. We also find 
evidence for a second emission line at 6.7--6.9~keV,
which we attempt to model by reflection from a highly ionised disc.
A `soft excess', evident as an upward curvature in the continuum 
flux below $\sim$1.5~keV, cannot be explained solely by enhanced reflection
from the ionised disc. The RGS spectrum shows only weak discrete
emission and absorption features in the soft X-ray spectrum,
supporting our conclusion that the soft excess emission in \mrk\ represents
the high energy portion of optically thick, 
thermal emission from the inner accretion disc.  
\end{abstract}


\section{Introduction}
In the standard paradigm the primary luminosity of an AGN originates in an 
accretion disc around a super-massive black hole. In the X-ray band, 
a hard power-law
component generally dominates above $\sim$2~keV in the well-studied
Seyfert~1 galaxies and is believed to arise in a hot corona above the 
surface of the accretion disc, where
optical/UV photons from the disc are Comptonised to X-ray energies.
These X-rays in turn illuminate the disc, being either `reflected'
towards the observer or thermalised back into
optical/UV emission (Mushotzky \et 1993). In the X-ray spectral band, 
evidence for this disc reflection component 
may be seen in the form of a broad fluorescent Fe K$\alpha$ line 
near 6.4~keV, an Fe K edge at $>7$~keV and a
Compton `hump' at $>10$~keV . All these features have been observed in
bright Seyfert~1 galaxies (e.g. Pounds \et 1990, Nandra \& Pounds 1994). 
The improved resolution of \asca\ showed the Fe K$\alpha$ line was 
often broad with excess flux, particularly in a `red' wing to the
line, explained by emission from the inner disc in a region of both high
velocities and high gravity (Tanaka \et 1995, Nandra \et 1997a).
  
The substantial gains in sensitivity, bandwidth and spectral
resolution of \xmm\
and \chandra\ are now beginning to further qualify this standard picture,
with indications that the energy, 
breadth and strength of the broad Fe K$\alpha$ line (the inner disc 
component) is a function of the 
luminosity and hardness of the X-ray continuum emission. Also, there
is growing evidence for a narrow emission line at $\sim$6.4~keV 
having a distinct origin, away from the inner disc 
(e.g., Reeves \et 2001, Yaqoob \et 2001, Kaspi \et 2001).

In this paper
we report on the \xmm\ observation of \mrk, the most luminous, hard
spectrum Seyfert 1 galaxy observed in the early phase of the mission.
At $z=0.0344$, \mrk\ has a typical X-ray luminosity (2--10~keV) of 
$3\times10^{44}$~erg s$^{-1}$. 
The Galactic absorption column towards \mrk\ is $4.4\times10^{20}$~cm$^{-2}$
(Murphy \et 1996), rendering it easily visible over the whole 
($\sim$0.25--10~keV)
spectral band of the EPIC detectors on \xmm.
First detected by \ariel\ (Cooke \et 1978), an \exosat\ observation
showed the X-ray spectrum of \mrk\ to be described by a power-law 
of index $\Gamma=1.9$ and a `soft excess' of $kT=70$~eV
(Singh \et 1989). \ginga\ observations in 1988 showed the 2--20~keV
spectrum could be fitted by a power-law of index of $\Gamma=1.6-1.7$
(steepening with increasing intensity), a variable soft excess with a
relatively high (black body) temperature of $kT=300$~eV, and a weak
emission line near 6 keV (Singh \et 1990). 
During a simultaneous \ginga\ and \rosat\ observation in 1990, \mrk\
was seen at a `typical' 2--10~keV flux of $5\times10^{-11}$~ergs~cm$^{-2}$
~s$^{-1}$, the (at that time) uniquely large bandwidth (0.15--15~keV)
yielding a complex continuum fit, of a power-law of index of $\Gamma=1.84$,
with some flattening at higher energies due to reflection, a soft
excess modelled by a black body of
$kT\sim80$~eV, and an `emission line' near 1 keV, possibly from Fe L 
(Pounds \et 1994, see also Turner \et 1991).
The total of 4 \ginga\ observations of \mrk\ also provided evidence 
for a weak Fe K emission line at $\sim$6.55~keV (Nandra \& Pounds 1994).

Fitting a narrow line to the 1994 \asca\ observation refined
the peak energy to $E=6.37\pm0.07$~keV, with an EW of $\sim$70~eV,
increasing to $\sim$200~eV for a broad line fit (Nandra \et 1997a).
Interestingly, the \asca\ spectral fit for \mrk\ indicated the presence
of a warm absorber rather than a `soft excess'
(Reynolds 1997,
George \et 1998). 

\section{Observations and Data reduction}

\mrk\ was observed on 2000 October 25 during orbit 161 of the PV 
phase of \xmm. The observations with the EPIC MOS (Turner \et 2001) and PN
(Str\"{u}der \et 2001) detectors were 
of $\sim$25~ksec and $\sim$27~ksec duration, respectively, all in 
Small Window mode. The simultaneous RGS observation lasted
$\sim$30~ksec.  
The EPIC data were screened with the XMM SAS (Science Analysis 
Software) and pre-processed using the CCD gain values for 
the time of the observation. 
X-ray events corresponding to patterns 0--12 for the 2 MOS cameras 
(similar to grades 0,2,3,4 in \asca) were used; for the PN, only 
pattern 0 events (single pixel events) were selected. 
A low energy cut of 200 eV was applied to the data 
and known hot or bad pixels were removed during screening. 
The non-X-ray background remained low throughout the observation. 

We initially extracted source and background spectra for 
the PN and both MOS
detectors separately. In each case a circular source region of
45\arcs\ radius was defined around the
centroid position of \mrk\, with the background being taken from an
offset position close to the source. 
As there was little variability of
\mrk\ ($<$10\%) during the observations and no significant 
difference between the MOS~1 and MOS~2 spectra, we combined these into a
single spectral file to maximise signal-to-noise. 
The integrated 
X-ray flux from \mrk\, over the 0.5--10 keV range, was 
$2.6\times10^{-11}$~ergs~cm$^{-2}$~s$^{-1}$. At this flux
level, the effect of photon pile-up in Small Window mode is
negligible.
The background-subtracted spectra for the PN and combined MOS detectors
were then fitted, using \textsc{xspec v11.0}, with the latest response
matrices produced by the EPIC team; the systematic level of uncertainty 
is $<5$\%.  Finally spectra were binned to a minimum of 20 counts per
bin, to facilitate use of the $\chi^2$ minimisation technique.
In the spectral analysis described in the following section, values 
of $ H_0 = 50 $~km\,s$^{-1}$\,Mpc$^{-1}$ and $ q_0 = 0.5 $ are 
assumed and all fit parameters are given in the AGN rest-frame.
Errors are quoted at
the 90\% confidence level (e.g. $\Delta \chi^{2}=2.7$ for one 
interesting parameter). 

\placefigure{fig1}

\section{Spectral Analysis}
 
\subsection{Fe K emission}

We initially fitted the hard X-ray (2--10 keV)
spectrum of \mrk\ for the PN and MOS data separately, with a power 
law and neutral absorption corresponding
to the line-of-sight Galactic
column of $N_{H}=4.4\times10^{20}\rm{cm}^{-2}$. 
For the PN data the best-fit power-law index of $\Gamma=1.64\pm0.01$
showed significant excess flux between 6--7 keV suggestive of 
iron K-shell emission. Very similar residuals were found from a 
fit to the MOS data, albeit with a slightly flatter power law index 
of $\Gamma=1.60\pm0.01$. Figure 1 shows the 
ratio of measured flux to
best-fit power law for both detector systems, with evidence for
a narrow emission line near 6.4 keV and additional emission to the `blue'
side. We then proceeded to fit the MOS and PN data simultaneously, 
but allowing for a small normalisation difference. The power law
indices were left free. 
The sequence of fits to the combined EPIC data for the 2--10 keV band
are summarised in Table 1. 

\placetable{tab1}

The initial comparison of the combined data set with a power law plus 
Galactic absorption
gave an unacceptable fit (fit 1 in table 1), mainly due to 
excess measured flux in
the 6--7~keV region. The addition of  
a Gaussian emission line, 
with energy and width free, produced a significantly improved fit 
($\Delta\chi^2=87$ for 
3 additional parameters; fit 2 in table 1). The 
line appears marginally resolved
(with intrinsic width $\sigma=80\pm35$~eV), has an EW of $\sim$82 eV 
and lies at 
$6.37\pm0.03$ keV, in close agreement with the narrow line fit to the
\asca\ data. We note this rest energy is consistent with fluorescence from 
neutral iron. The addition of a second narrow Gaussian 
emission line gave a  
further improvement 
($\Delta\chi^2=17$ for 3 extra parameters), with the overall 
fit-statistic now $\chi_{\nu}^2=1386/1288$ (fit 3 in table 1). 
This second emission line is less well determined, but clearly lies to
the high energy side of the first line, with a best-fit energy of
$E=6.91\pm0.09$~keV and EW $\sim$44~eV.
Inclusion of the 2 Gaussian lines in the fit increased the power law index by
$\sim$0.03, to a value consistent with the directly measured index
from 2--5 keV.

\subsection{The Soft Excess}

\placefigure{fig2}

Extrapolating the best-fit 2--10 keV spectrum (fit 3 in table 1) 
to the lower useful limit of EPIC ($\sim$~0.25 keV)
revealed a clear excess of soft X-ray flux below $\sim2$~keV (figure 2). 
Modelling this `soft excess' with a second power law, or a single
black body,
gave a poor fit, the breadth of the soft excess
requiring 3 black-body components to approach a minimum value of
chi-squared.
Table 2 (line 2) details this best-fit to the overall (0.25--10 keV)
spectrum. 
Parameterised in this way, the multiple BB components account for 
$\sim$40\% of the 
de-reddened luminosity of \mrk\ of 
$1.5\times10^{44}$~erg s$^{-1}$ between 0.25--2~keV, but a negligible
fraction of the harder (2--10 keV) luminosity of 
$1.4\times10^{44}$~erg s$^{-1}$. 

The RGS spectrum showed no strong
features between 6--38 \AA (0.3--2~keV), other than the absorption edge 
of neutral interstellar oxygen at 23.3 \AA (0.53 keV), supporting the 
EPIC modelling of
the soft excess in terms of thermal continuum emission.
However, detailed examination did 
reveal a weak absorption trough centred at 16.1 \AA, 
and an unresolved absorption line at $13.52\pm0.05$~\AA, 
which we
tentatively identify with a blend of Ne IX and Fe XIX (Figure 3a). Both
features are weak, with equivalent widths 
of
$3.8\pm1.8$~eV and
$2.2\pm1.0$~eV,
respectively. Although the long wavelength `edge' of the `trough',
at $16.4\pm0.2$~\AA,
is close to the K-edge of OVII (16.8\AA),
the shape is unlike that of continuum absorption and we favour
identification with absorption via an unresolved transition array 
(mainly 2p--3d transitions) in Fe M-shell ions, as found  
in the RGS spectrum of the infrared-loud quasar IRAS
13349+2438 (Sako \et 2001; see also Chenais-Popovic \et 2000).
  
The only emission features we find are consistent with the triplet of
helium-like OVII (see Figure 3b). The equivalent widths
of the individual components are $0.5\pm0.5$~eV (R), $1.5\pm0.7$~eV (I)
and $0.6\pm0.4$~eV (F).
The 
relative weakness of 
the resonance
line is consistent with emission from a photo-ionised gas, a point we
take up again in section 5. For now, we reiterate our main finding, that 
the high resolution RGS spectrum supports the 
EPIC modelling of
the soft excess of \mrk\ in terms of thermal 
continuum emission. 

\placefigure{fig3}

\section{Modelling the Spectrum}

The \xmm\ spectrum of \mrk\ is strikingly similar 
to that of the
low luminosity quasar, Mkn 205 (Reeves \et 2001), with the main
spectral features in common being: a narrow Fe K$\alpha$ line at
$\sim$6.4 keV,
additional emission centred at $\sim$6.7--6.9 keV, and a moderately strong,
featureless soft excess below $\sim$2 keV. On a more detailed comparison,
the underlying power law in \mrk\ is only slightly harder ($\sim$0.1 in
index), and the soft excess is very similar 
(measured by the ratio of BB to PL flux at 
0.25--2~keV: $\sim$0.7 in \mrk\ and $\sim$0.6 in Mkn 205).  

\placefigure{fig4}

\placetable{tab2}

In order to test our parameterisation of the \mrk\ spectrum against a
physical model, we have compared our \xmm\ data for \mrk, and for the
similar case of Mkn 205, with a model including the effects of X-ray
reflection from an ionised disc. Since the narrow line at 6.4 keV is
believed to be unrelated to the disc emission it is excluded from the
modelling. The model used here will be described
in detail in Nayakshin et al (2001, in preparation), and here we only 
summarise
it. The approach is similar to the one used for stellar
spectra. Specifically, a large number of models are computed for a range of
values of the spectral index, the incident X-ray flux, disc gravity,
the thermal disc flux and iron abundance. The output of each model is an
un-smeared reflected spectrum for 5 different inclination angles
ranging from nearly pole-on to nearly face on, stored in a look-up
table. The `lamppost' geometry is then assumed, with free parameters of
the model being the height of the X-ray source above the disc, $h_X$,
the dimensionless accretion rate through the disc, $\dot{m}$, the
luminosity of the X-ray source, $L_X$, the inner and outer disc radii,
and the spectral index. This defines the gravity parameter, the ratio
of X-ray to thermal fluxes, etc., for each radius, which enables the 
look-up table to be used to approximate the reflected spectrum. This
procedure is repeated for about 30 different radii. The total disc
spectrum is then obtained by integrating over the disc surface,
including relativistic smearing of the spectrum for a non-rotating
black hole (e.g., Fabian et al. 1989).

In using this model, the \xmm\ data were first fitted between 2--10~keV, 
excluding the
narrow $\sim$6.4~keV line, and freezing $h_X$ at $10 R_S$. Good fits
were obtained for both \mrk\ (fit 4 in table 1) and Mkn 205, with the Fe K 
feature being
identified with predominantly helium-like Fe, emitted at 6.7~keV, and
observed at an inclination of $\sim$30~degrees.  Figure 4 illustrates
how the model is constrained by the overall profile, as well as the 
energy and strength of the line.  
We note in passing that the underlying power-law index is now $\sim$0.1 
higher as the fit also models
the flattening due to continuum reflection at high energies.  
The `red wing' of the 6.7 keV line constrains the inner disc radius 
for \mrk\ to
be $R_{\rm in}<10 R_S$.
While not tightly constrained, the remaining parameters in the ionised
disc model fit for \mrk\ are interesting, with a best fit for
$L_X/L_d$ $\sim$1 and an accretion rate $\dot{m}=0.15$ (where
$\dot{m}=1$ corresponds to the Eddington limit).
 
It is important to
note, however, that the spectral table employed in the fits assumed
that the power-law continuum extends out to $E_c = 200$ keV, whereas
a recent BeppoSAX observation of \mrk\ has suggested that the high energy
spectrum is modified by an approximately exponential roll-over at
$\sim 75$ keV (Perola \et 2000). If confirmed, this would be significant 
for our present fits 
since,
for hard X-ray spectra such as \mrk, $E_c$ determines the
Compton temperature of the gas, and consequently the ionisation state
of the reflector (see the appendix and figures in Nayakshin \&
Kazanas 2001a). To test this we re-ran the code of Nayakshin et
al. (2000) with the best fit parameters described above,
but with a cutoff energy of $E_c = 75$ keV. We indeed found that this
led to a substantially cooler temperature and a higher
fraction of He-like iron in the skin. The latter then produced a much
stronger He-like line than required by our data ($\sim$300~eV). 
To recover the 
ionization state of the skin to the degree corresponding to our $E_c =
200$ keV fit, we needed to increase $L_X/L_d$ to a value much higher
than unity.
This contrasts sharply with a deduction from the SED
of \mrk\ which 
suggests that $L_X/L_d$ is much {\em smaller} than unity (see Walter
\& Fink 1993). In addition, Kriss et
al. (2000) find that the ionising UV luminosity of \mrk\ is $\sim
3\times 10^{45}$ erg s$^{-1}$, while its 2-10 keV luminosity is
$L_{2-10}\sim 3\times 10^{44}$. Clearly, for the condition $F_{X} >>
F_{\rm disc}$ (required by our  
photo-ionisation model) to be consistent with the observed luminosity
ratio, we need
to invoke a geometry of the X-ray source different from the `lamppost'. 
An origin of the hard X-ray emission in magnetic flares above the disc 
is an attractive 
possibility, since in that case only a small fraction of the disc is
illuminated by the flare and hence the local ionising flux is much
larger than $L_X/4\pi R^2$ (for estimates see Nayakshin \& Kazanas
2001b). In addition, the hard X-ray spectrum of \mrk\ puts
constraints on the amount of cooling due to reprocessed radiation,
again requiring a patchy corona (Haardt, Maraschi
\& Ghisellini 1994).

An important property of all ionised discs is that they reflect
strongly at low X-ray energies where neutral discs absorb the
internally scattered photons. To check the contribution of such
enhanced reflection on the \xmm\ spectrum, the best-fit ionised disc
model at 2--10~keV (fit 4 in table 1) was extrapolated over the full
data bandwidth. A 'soft excess' is still strongly required, but now the
hottest BB component is removed by the enhanced reflection of the 
ionised disc below $\sim$2~keV. This full-band fit
is summarised in line 3 of table 2.

A by-product of the ionised disc fit is that the narrow
6.4~keV line EW falls to $\sim$50 eV, and is now completely unresolved
by the EPIC detectors ($\sigma<75$~eV), but still remains
statistically very significant.

\placefigure{fig5}

\section {Discussion}

The first \xmm\ observation of \mrk\ found the source in a relatively
faint state, with the 2--10~keV flux about half of its previously measured
value. That may account for the mean power-law being somewhat
flatter than in earlier measurements.
However, our
spectral fits clearly show the presence of a `soft excess' which lies
well above the power-law even when the latter is enhanced at low energies 
by reflection from a strongly ionised disc. Of particular interest is
the appearance of two components in the Fe K band, and we note the
similarity to the \xmm\ observation of Mkn 205.   

The most prominent feature, at
6.4~keV, we identify with the narrow line component now
emerging as a common feature of Seyfert 1 spectra (e.g. Reeves \et 2001, 
Yaqoob \et 2001, Kaspi \et 2001). Recognition of this
separate component to the Fe K emission has two important
consequences. First, fitting of the broader line profiles commonly seen
in \asca\ spectra of bright (generally lower luminosity) Seyferts 
should now allow for a separate
contribution to the line core at 6.4 keV.
Nandra \et (1997a), in their survey of \asca\ spectra of AGN, found no
requirement for a narrow Fe K$\alpha$ line, but with upper limits to the EW 
of order $\sim$100~eV. However, Lubinski \& Zdziarski (2001)
in their analysis of a larger sample of Seyfert 1s did suggest that the narrow
emission line component may be important.
Secondly, the detection of a strong, narrow iron emission line at 
6.4 keV indicates that a substantial quantity of cool 
reprocessing material --- lying outside of the line-of-sight
since the X-ray spectrum shows no intrinsic  
absorption --- is illuminated by the central X-ray source. 
The line strength (EW$\sim50-75$~eV) implies 
that the cool matter subtends a solid angle 
of at least 1 $\pi$ steradian, assuming it is Thomson thick
and has a solar abundance of iron (e.g. George \& Fabian 1991).  
The most likely location of such material, within the framework of current
AGN models, would appear to be the molecular torus 
(Antonucci 1993), with 
hard X-rays from the central engine being reflected off the 
inner surface of the torus and into the line-of-sight of the observer
(Krolik, Madau \& Zycki 1994, Ghisellini \et 1994). 
We
cannot, however, at present rule out other alternatives, given the EPIC CCD
spectral resolution, including emission from BLR clouds
(Leahy \& Creighton 1993).

The detection of emission from strongly ionised Fe is of particular
interest in relation to exploring the inner accretion disc. 
We note that high ionisation Fe K lines have previously been suggested
by \asca\ data of several radio-quiet quasars (Reeves \& Turner 2000),
while the so-called X-ray Baldwin effect has been invoked to describe
the apparent tendency (in \asca\ spectra) for the Fe K emission to be
weaker for higher luminosity AGN (Nandra \et 1997b). The \xmm\
observations of \mrk\ and Mkn 205 now confirm, for 2 high luminosity AGN,
that the Fe-K emission is indeed relatively weak.
A reduction in strength of the observed He-like Fe
K line can be understood if the accretion disc develops an
over-ionised hot skin produced by the strong hard X-ray illumination
(Nayakshin \et 2000, Ballantyne et al. 2001). 
Assuming the low value for the high energy spectral rollover as observed
by BeppoSAX, we find that a large ratio of the ionising flux to the
local disc thermal flux is required to keep the `skin' hot
enough. Given the observed ratio of X-ray luminosity to
optical-UV luminosity $\sim 0.1$, this then requires a flux
concentration which is only plausible if the X-rays originate from
localised sources, e.g.
magnetic flares near the disc surface. 

Aside from the Fe K line emission, the EPIC X-ray spectrum of \mrk\
appears to be devoid of other discrete emission or absorption
features. In particular the absence of strong recombination emission 
from Si, Mg, Ne and O constrains the disc reflection model
to high values of the effective ionisation parameter. 
Thus we are apparently observing the bare nuclear continuum emission
in \mrk, with the soft excess flux 
arising as thermal emission from the
inner accretion disc.  

Finally, we comment briefly on the RGS
spectrum. The detection of an absorption
trough at 16--17 \AA, which we tentatively identify with absorption in
Fe M ions, is interesting, particularly in comparison with the
discovery of this feature in IRAS 13349+2438 from an earlier \xmm\ 
observation (Sako \et 2001). In the latter case the absorption
was interpreted as arising in a cool gas component associated with the
dust that causes strong reddening in the optical band (E(B-V)=0.3).
In \mrk\ the optical reddening is more than an order of magnitude less,
implying that here the temperature of this 'cool' material remains too
high for dust grains to survive. A further implication of the Fe M
absorption in \mrk, if confirmed, is that it raises a question about the
common occurrence of OVII absorption edges, particularly in the
analysis of \asca\ data (e.g. Reynolds 1997, George \et 1998).
The narrow absorption line at 13.5 \AA, and the emission from OVII
require a zone of higher ionisation in the line-of-sight material, 
and the ratio of components in
the OVII triplet suggest an origin in a photoionised gas with electron
density greater than $10^{11}$~cm$^{-3}$ and electron temperature
less than $10^{6}$~K (Porquet and Dubau 2000).
   
\section{Conclusions}

Observations with \xmm\ have revealed 2 apparently separate
components to the Fe K line
emission of the high luminosity Seyfert 1 galaxy,
\mrk. Instead of a broad iron line component centred near 6.4 keV,
with a strong `red' wing, as found in lower luminosity Seyfert galaxies,
and explained by reflection from ``cold'' matter in 
the inner accretion disc, we find a relatively weak line 
centred at
6.7--6.9 keV.  We model this feature in terms of
fluorescence
from He-like iron in the highly ionised inner disc material. The energy,
equivalent width and profile of this feature are all important in the
model fitting, as is the high energy cutoff of the illuminating X-rays.
If a cutoff as low as $E_c = 75$ keV (Perola \et 2000) is confirmed
then our modelling suggests the illuminating X-ray flux $F_{X} >> 
F_{\rm disc}$ in order that the 
He-like line is not much stronger than we 
observe. Our tentative conclusion is to suggest a geometry in which 
the hard X-ray emission arises in magnetic flares above the disc (Nayakshin 
\& Kazanas 2001b).

The narrow line at 6.4
keV appears to be an increasingly common feature in the higher
quality spectral data emerging from \chandra\ and \xmm, and probably 
arises from Compton scattering off distant, cool matter. 
If this material is in the putative
molecular torus, then --- as with Mkn 205 --- this provides important
evidence that a substantial torus exists for AGN with bolometric 
luminosities as high as $10^{45}$~erg s$^{-1}$.
However, it appears that narrow 6.4
keV iron lines are weak or absent in
higher luminosity QSOs. For example, upper limits of 10 eV have been set
on narrow 6.4 keV lines in the high luminosity quasars 
PKS0558$-$504 (O'Brien \et 2001), S5 0836+710 
and PKS 2149$-$306 (Fang \et 2001). Whatever the origin of the neutral gas, the
observations imply a smaller covering factor for such gas in
high-luminosity objects. If confirmed by observations of a larger
sample of AGN, the apparent correlation of narrow 6.4 keV line
strength with bolometric luminosity is consistent with
torus models involving a dusty, disc-driven
hydromagnetic wind (Konigl \et 1994),
which predict that the dust distribution will
be `flattened' due to radiation pressure in high luminosity AGN, hence
reducing the solid angle subtended by the torus.

\acknowledgements
This work is based on observations with \xmm, an ESA science 
mission with instruments and contributions directly funded by 
ESA Member States and NASA. 
The authors would especially like
to thank the EPIC and RGS instrument teams, and the SOC and SSC staff
for making the
observation and subsequent analysis possible. JR is supported
by a Leverhulme Research Fellowship and 
KP by a Research Studentship from the UK Particle Physics and Astronomy
Research Council.

\pagebreak

\begin{deluxetable}{lcccccccc}
\tablecolumns{9}
\tablewidth{0pc}
\tablecaption{Spectral fits to joint PN and MOS data over 2--10 keV
energy band}
\tablehead{
\colhead{Fit} & 
\colhead{$\Gamma$} & 
\multicolumn{3}{c}{1st line} &
\multicolumn{3}{c}{2nd line} & 
\colhead{$\chi^{2}$/dof}\\
\colhead{} & 
\colhead{} & 
\colhead{E\tablenotemark{a}} & 
\colhead{$\sigma$\tablenotemark{b}} & 
\colhead{EW\tablenotemark{c}} & 
\colhead{E\tablenotemark{a}} & 
\colhead{$\sigma$\tablenotemark{b}} & 
\colhead{EW\tablenotemark{c}} & 
}
\startdata
1. P-L only & 1.64$\pm$0.01 & & & & & & & 1490/1294 \\
2. P-L + Gauss & 1.66$\pm$0.02 & 6.37$\pm$0.03 & 80$^{+40}_{-34}$ &
82$^{+19}_{-17}$ & & & & 1403/1291 \\
3. P-L + 2$\times$Gauss & 1.67$\pm$0.02 & 6.36$\pm$0.03 & 82$^{+38}_{-33}$ &
85$^{+19}_{-18}$ & 6.91$\pm$0.09 & 120$^{+120}_{-78}$ &
44$^{+24}_{-20}$ & 1386/1288 \\
4. Xion\tablenotemark{d} + Gauss\tablenotemark{e} & 1.76$\pm$0.05 & 6.36$\pm$0.04 & $<75$ &
47$^{+17}_{-16}$ & \ & \ & \ & 864.0/874 \\
\enddata
\tablenotetext{a}{Rest energy of line (keV).}
\tablenotetext{b}{Intrinsic (1 sigma) width of line (eV).}
\tablenotetext{c}{Equivalent width of line (eV).}
\tablenotetext{d}{Xion - Ionised
reflection model (Nayakshin \et 2000) --- see text. Model best-fit
parameters are L$_{X}$/L$_{d}=1$, accretion rate
$\dot{m}=0.15m_{edd}$, inclination$=30$\degg, and inner radius
$r_{in}=3r_{s}$.}
\tablenotetext{e}{Fit to PN data only.}
\end{deluxetable}

\begin{deluxetable}{lccccc}
\tablecolumns{6}
\tablewidth{0pc}
\tablecaption{Broad-band (0.25-10 keV) continuum spectral fits to joint
PN and MOS data.}
\tablehead{
\colhead{Fit\tablenotemark{a}} & 
\colhead{$\Gamma_{h}$} & 
\colhead{$kT_{1}$\tablenotemark{b}} & 
\colhead{$kT_{2}$\tablenotemark{b}} & 
\colhead{$kT_{3}$\tablenotemark{b}} & 
\colhead{$\chi^{2}$/dof}
}
\startdata
1. P-L only & 1.99$\pm0.01$ & & & & 4964/1704 \\
2. P-L + 3$\times$bbody & 1.62$\pm$0.02 & 18$\pm$5 & 107$\pm$3 &
332$\pm$12 & 1834/1698 \\
3. Xion\tablenotemark{c} + bbody\tablenotemark{d}
& 1.75$\pm$0.02 & \ & 98$\pm$2 & \ &
1226/1183 \\
\enddata
\tablenotetext{a}{The region near 0.5 keV, where the neutral oxygen
absorption in the oxide layer of the EPIC detectors is not well
modelled at present, 
is excluded from each fit.}
\tablenotetext{b}{Temperature (kT) of blackbody emission (units eV).}
\tablenotetext{c}{Xion = Fit to ionised reflection model
(Nayakshin \et 2000) - see text for details.}
\tablenotetext{d}{Fit to PN data only.}
\end{deluxetable}

\begin{figure}
\begin{center}
\rotatebox{-90}{\includegraphics[width=12cm]{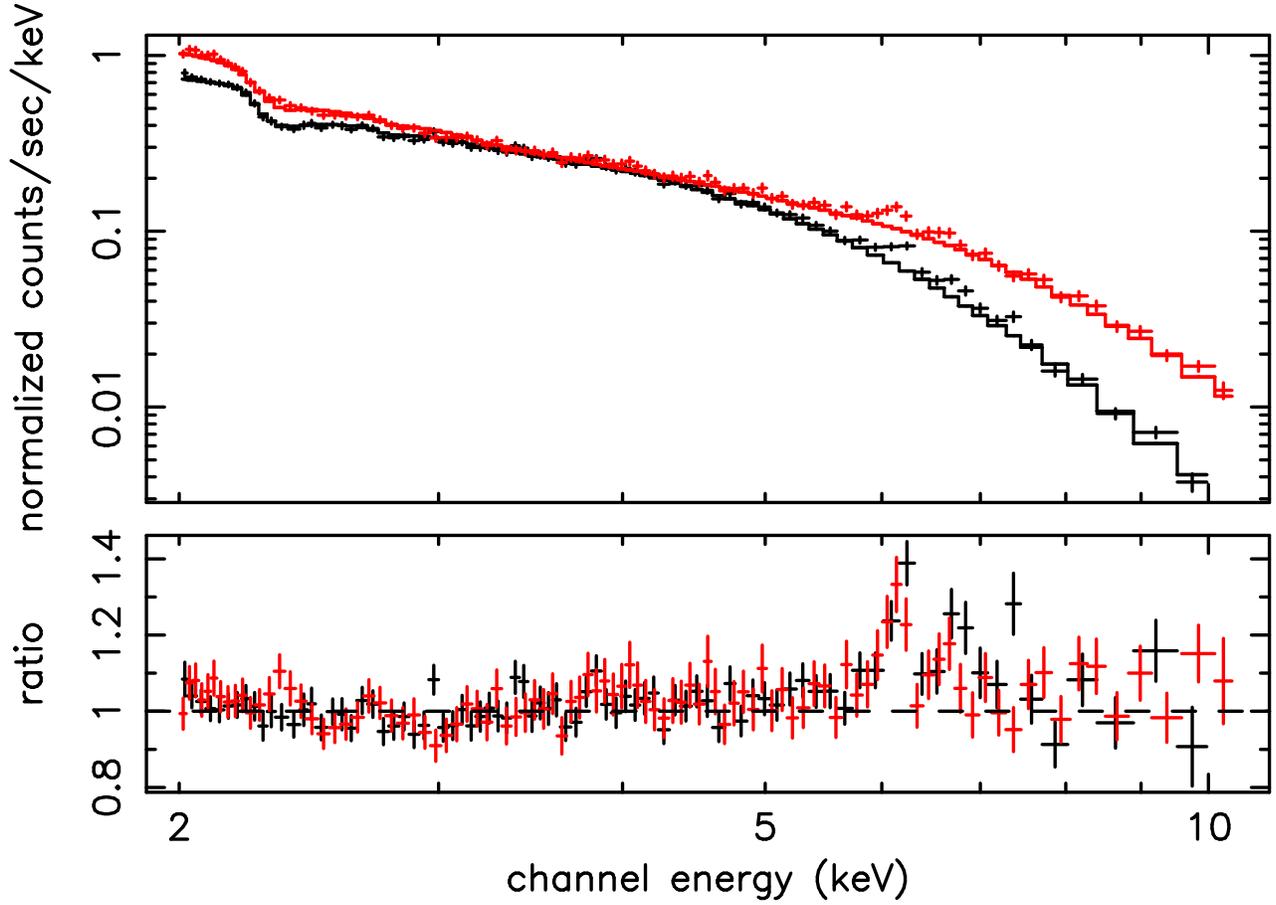}}
\end{center}
\caption{The 2-10 keV EPIC spectrum of Mkn 509, fitted with a power-law 
plus Galactic absorption. In this and subsequent figures data points
are shown as crosses and the fitted model as a full line. Here the 
MOS data are 
plotted in 
black, PN in red. Positive residuals are present in the iron K band,
corresponding to a narrow emission line at 6.4 keV 
(rest frame), and a second, weaker line at 6.7-6.9 keV.The lower
section shows the ratio of data to power law model.}
\end{figure}

\begin{figure}
\begin{center}
\rotatebox{-90}{\includegraphics[width=12cm]{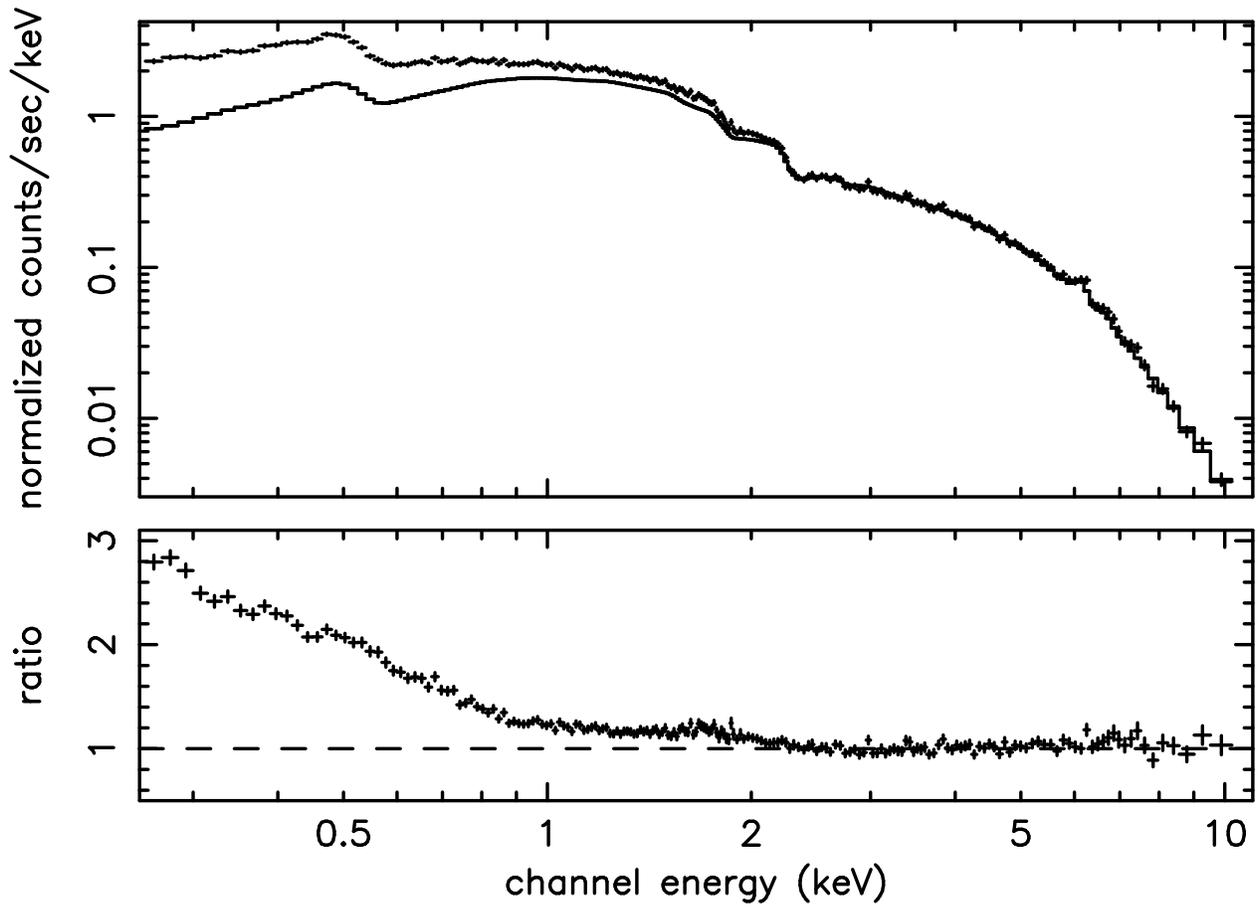}}
\end{center}
\caption{Comparison of data to extrapolation of the best-fit 
2-10 keV EPIC-MOS spectrum (fit 3, table
1) down to 0.25 keV. A clear `soft excess' is present at energies
below $\sim$2 keV.}
\end{figure}

\begin{figure}
\begin{center}
\rotatebox{-90}{\includegraphics[width=9cm]{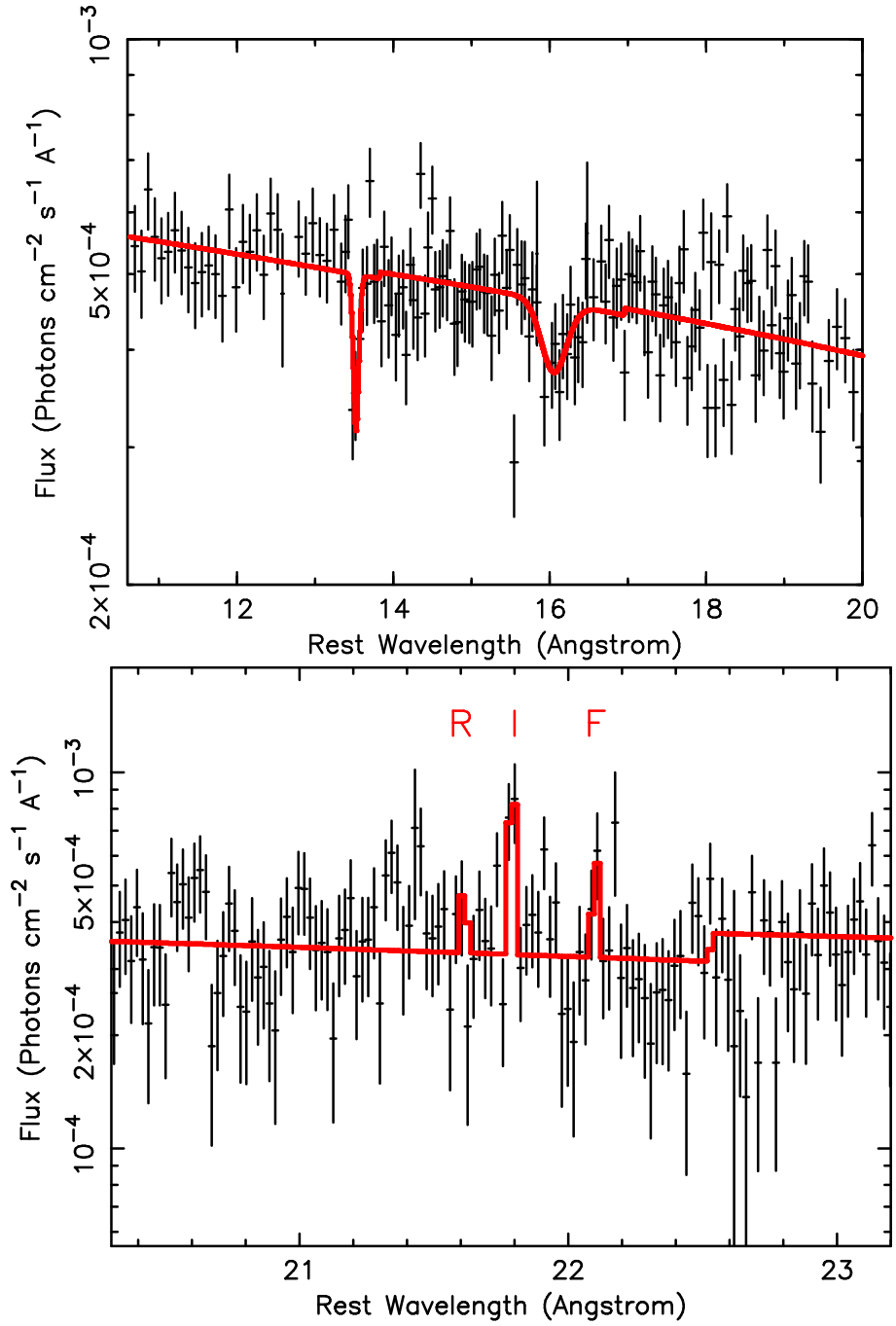}}
\\
\rotatebox{-90}{\includegraphics[width=9cm]{f3b.eps}}
\end{center}
\caption{(a) Sections of the RGS spectrum of Mkn 509 showing (a)
evidence for absorption due to M shell Fe and a blend of Ne IX and Fe XIX, (b) 
the emission 
triplet of OVII (see text for details).}
\end{figure}

\begin{figure}
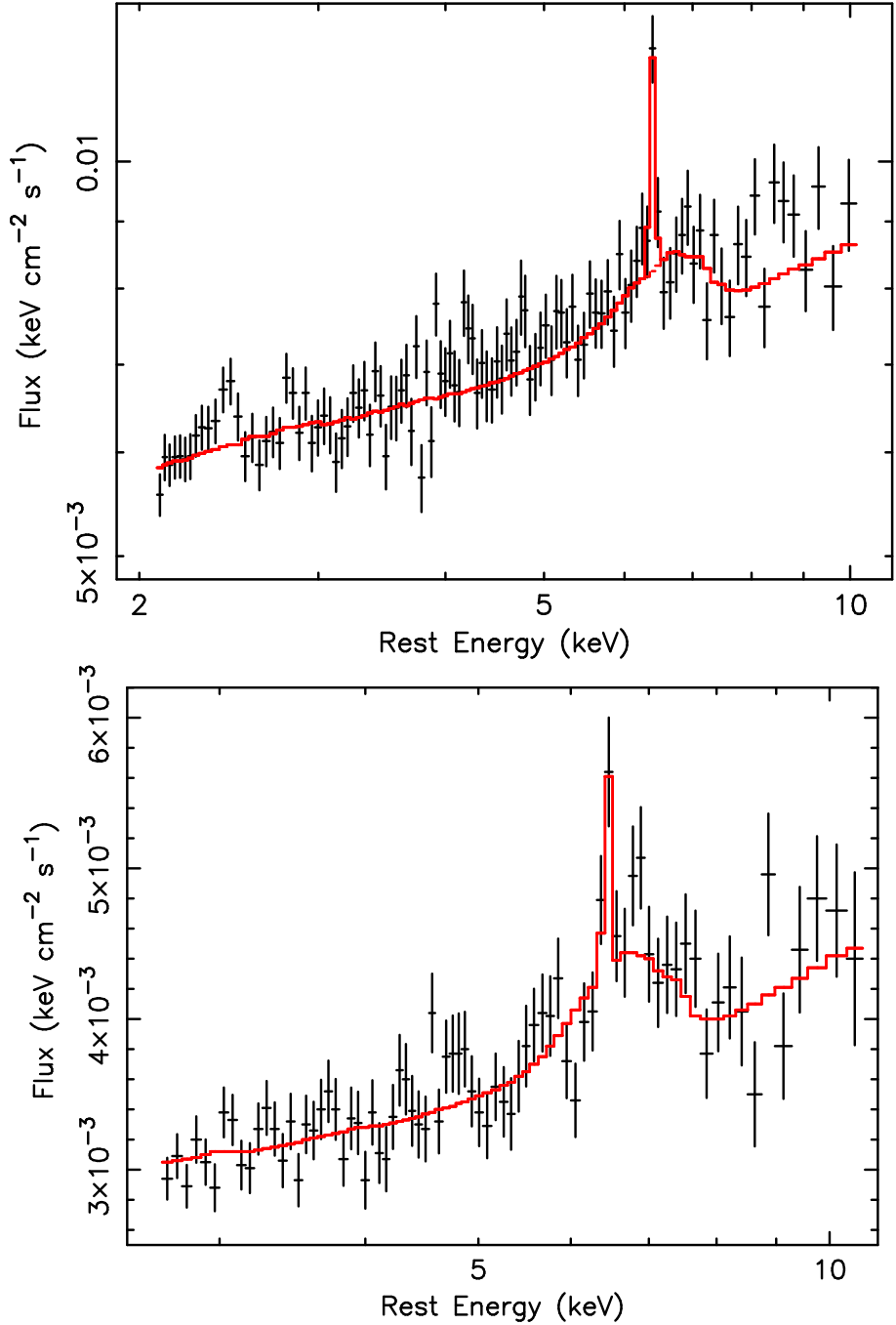

\begin{center}
\rotatebox{-90}{\includegraphics[width=9cm]{f4a.eps}}
\\ 
\rotatebox{-90}{\includegraphics[width=9cm]{f4b.eps}}
\end{center}
\caption{(a) The best-fit ionised reflection spectrum (2-10 keV band)
of Mkn 509, modelled using the code of Nayakshin \et (2000). 
X-ray reflection off an ionised disc can
account for the emission line peaking near 6.9 keV. (b) For
comparison, the EPIC spectrum of Mkn 205 (see Reeves \et 2001), modelled
with an ionised disc reflector as above. 
A broad {\it ionised} iron K$\alpha$ line is present
in both of these objects.}
\end{figure}

\begin{figure}
\begin{center}
\rotatebox{-90}{\includegraphics[width=12cm]{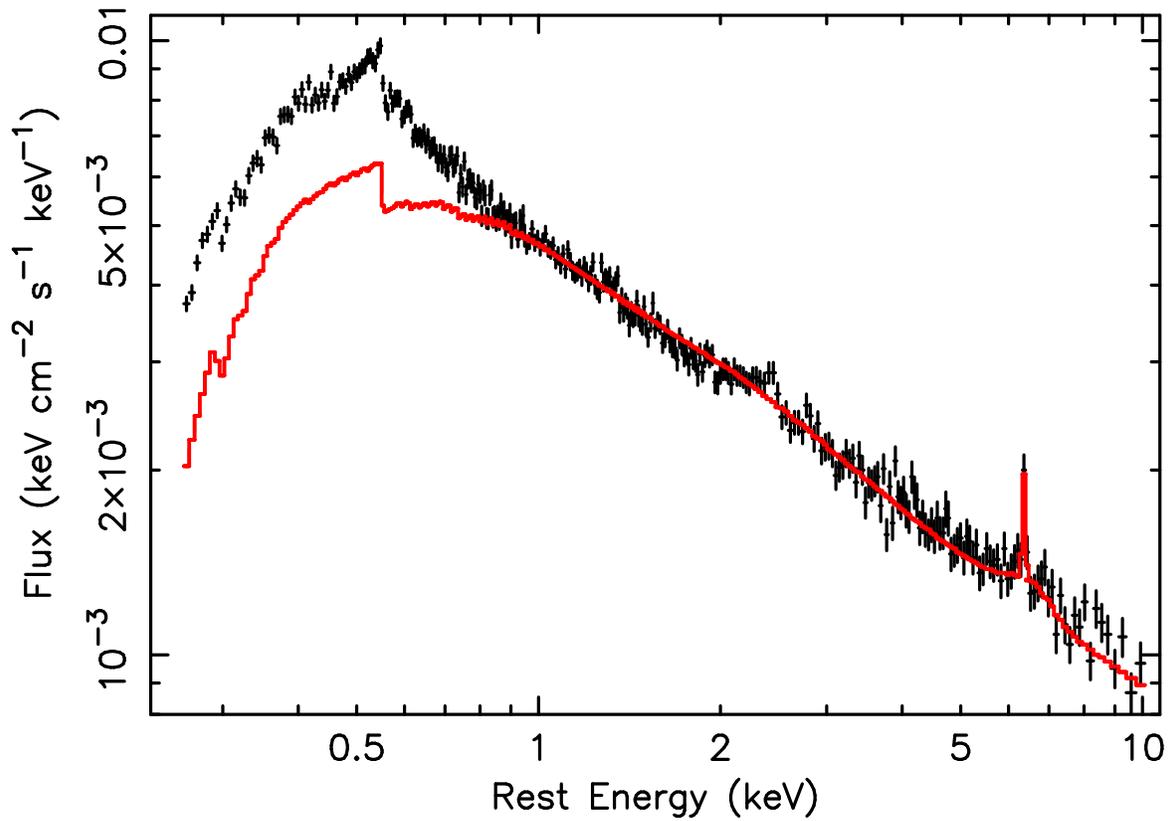}}
\end{center}
\caption{Extrapolation of the best fit disc reflection spectrum (fit
4, table 1) down to 0.25 keV. The model is now a good
description of the continuum down to $\sim$0.9 keV, below which
there remains a 
substantial excess of soft flux.}
\end{figure}

\end{document}